\pdfoutput=1

\documentclass[twoside,slac_one]{revtex4}
\usepackage{graphicx}
\usepackage{fancyhdr}
\usepackage{amsmath} % American Mathematics Society standards
\usepackage{bm}% bold math
\usepackage{amsxtra}
\usepackage{amssymb}
\usepackage{amsthm}
\usepackage{latexsym}
\usepackage{lscape}

\begin{document}

%Title of paper
\title{Studies of VCSEL Failures in the Optical Readout Systems of the ATLAS Silicon Trackers and Liquid Argon Calorimeters}

\author{Mark S. Cooke}
\affiliation{Lawrence Berkeley National Laboratory, Berkeley, CA, USA}

\begin{abstract}
The readout systems for the ATLAS silicon trackers and liquid argon calorimeters utilize vertical-cavity surface-emitting laser diodes to communicate between on and off detector readout components.  A number of these VCSEL devices have failed well before their expected lifetime.  We summarize the failure history and present what has been learned thus far about failure mechanisms and the dependence of the lifetime on environmental conditions.
\end{abstract}

%\maketitle must follow title, authors, abstract
\maketitle

\thispagestyle{fancy}

%%%%%%%%%%%%%%%%%%%%%%%%%%%%%%%%%%
\section{Introduction}

The ATLAS experiment~\cite{ATLASpaper} employs Vertical-Cavity Surface-Emitting Lasers (VCSELs)~\cite{VCSELbasics} in the optical readout systems of the silicon tracking detectors and liquid argon (LAr) calorimeters.  A number of these devices have failed well before their expected lifetime.  After introducing the basic features of a typical device and how the devices are used in the optical readout of each sub-detector, the history of the failures is recounted.  Results from a microscopic analysis of a failed device are presented next, followed by a discussion of measurements monitoring changes in the optical spectrum.  The last set of measurements presented study the lifetime of devices operated under extreme environmental conditions.  Contingency plans which are currently under way are then described, followed by concluding remarks. 

\section{VCSEL Basics}

\begin{figure}[bp]
\centering
\includegraphics[scale=0.45]{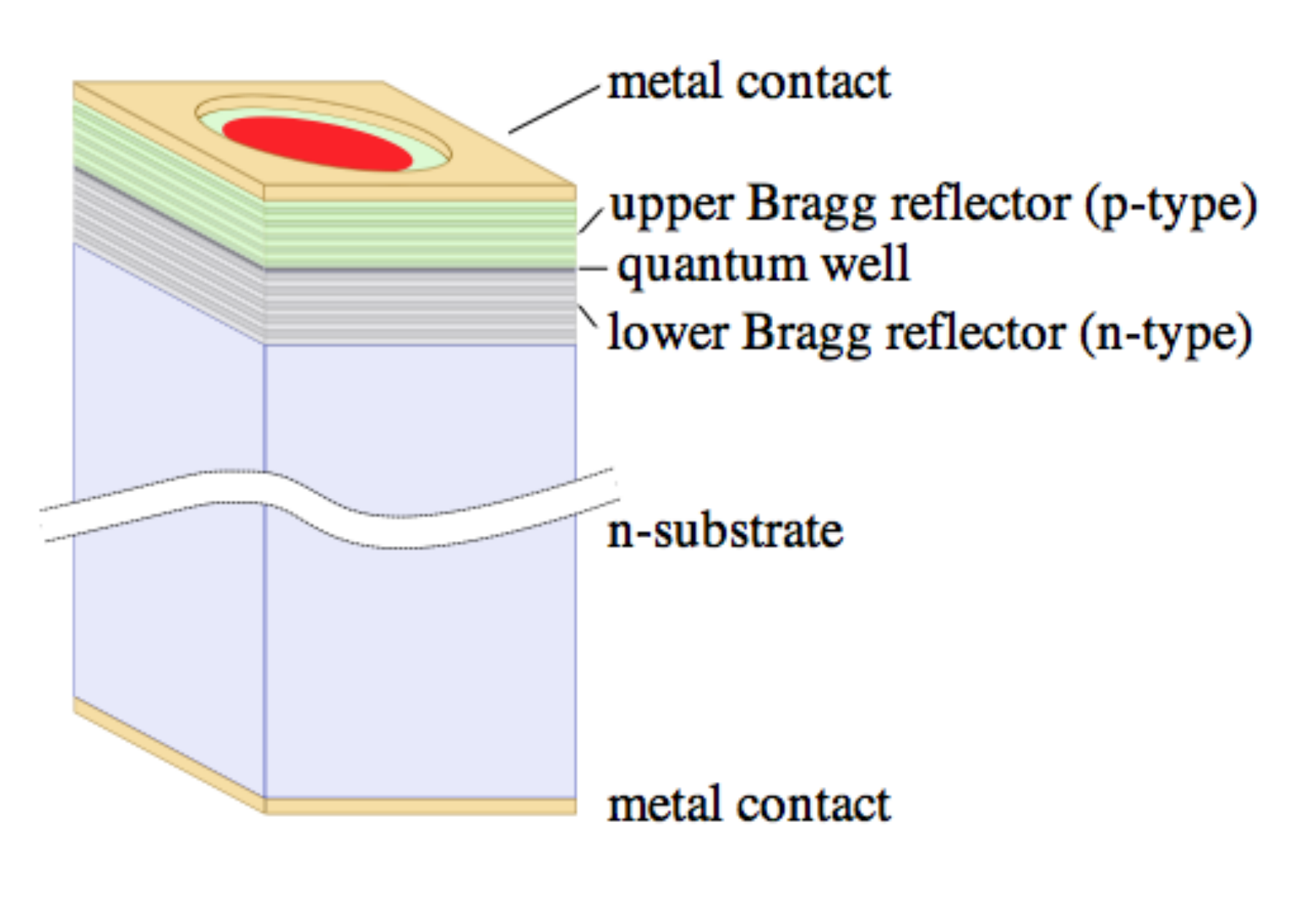}
\caption{ A schematic depiction a typical VCSEL device.}
\label{fig:VCSELschem}
\end{figure}

Figure~\ref{fig:VCSELschem} shows a schematic depiction of a typical VCSEL structure.  The laser resonator consists of two doped reflector regions about an active region.  The reflector regions alternate layers of high and low index of refraction (e.g. GaAs/AlGaAs).  Quantum wells in the active region provide optical gain, and enhanced transverse current confinement is typically achieved with an oxidized layer or a layer containing ion implants (not depicted in the figure).  VCSELs may be fabricated as single channel devices or in arrays.  Single channel VCSELs are often hermetically packaged, as the devices are known to be sensitive to damage from external sources~\cite{EXTref}, such as humidity~\cite{HUMref} and electrostatic discharge (ESD)~\cite{ESDref}.  Hermetic sealing of arrays is generally not an economic option, and some ATLAS VCSEL arrays have been coated with epoxy instead.

% -- VCSEL Figure --
% ------------------------

%%%%%%%%%%%%%%%%%%%%%%%%%%%%%%%%%%
\section{The Optical Readout of the ATLAS LAr Calorimeters and Silicon Trackers}

The front-end boards (FEBs)~\cite{FEBpaper} of the LAr calorimeters amplify, shape, digitize, and gain select the ionization signals.  In total, 1524 FEBs are distributed across 58 crates mounted on the cryostat walls of the calorimeters.  Each FEB is instrumented with an optical transmitter package (OTx) housing a single channel oxide VCSEL.  The FEB creates a serial data stream containing the digitized ionization pulses in triggered events.  The OTx transmits this stream across $\sim70$~m length optical fibers to back-end electronics located in an off-detector counting room.

The front-end modules of the pixel detector~\cite{PIXpaper} electrically transmit the address and time of pixels measuring charge over a predefined threshold in triggered events to opto-boards~\cite{PIXreadout} mounted on service panels about 1~m from the detector.  A total of 272 opto-boards are used to provide optical links to 6 or 7 pixel modules.  Oxide VCSEL arrays on the opto-boards transmit the data stream over $\sim70$~m fibers to the off-detector readout system.  In addition, so-called TX-plugins in the off-detector readout system contain oxide VCSEL arrays which similarly transmit trigger, timing, and control commands to the front-end modules.

The optical readout of the semiconductor tracker (SCT)~\cite{SCTreadout} and the pixel detector are similar, as the architecture of the latter was inherited from the former, with certain modifications to adapt to differing data rates, modularity, and radiation tolerances.  The most important differences for the present discussion are the use of proton implant single channel VCSELs on the on-detector opto-electronics of the SCT, and roughly a factor of four more SCT links compared to pixel links.  The off-detector readout electronics are nearly identical.

%%%%%%%%%%%%%%%%%%%%%%%%%%%%%%%%%%
\section{Failure History of ATLAS VCSELs}

Since January 2007, a total of 46 of the 1524 LAr OTx have failed~\cite{LARproc}.  The failure rate was most pronounced in 2009 and has markedly decreased since.  Access to the FEBs for replacement is only possible during long LHC shutdown periods, as it requires retracting large components of the ATLAS detector.  Two replacement campaigns have been carried out.  The first occurred in the Spring of 2009 and the second in the early months of 2011.  There are presently no dead OTx on the LAr calorimeters.
  
\begin{figure}[bp]
\begin{minipage}[b]{0.5\linewidth}
\centering
\includegraphics[scale=0.46]{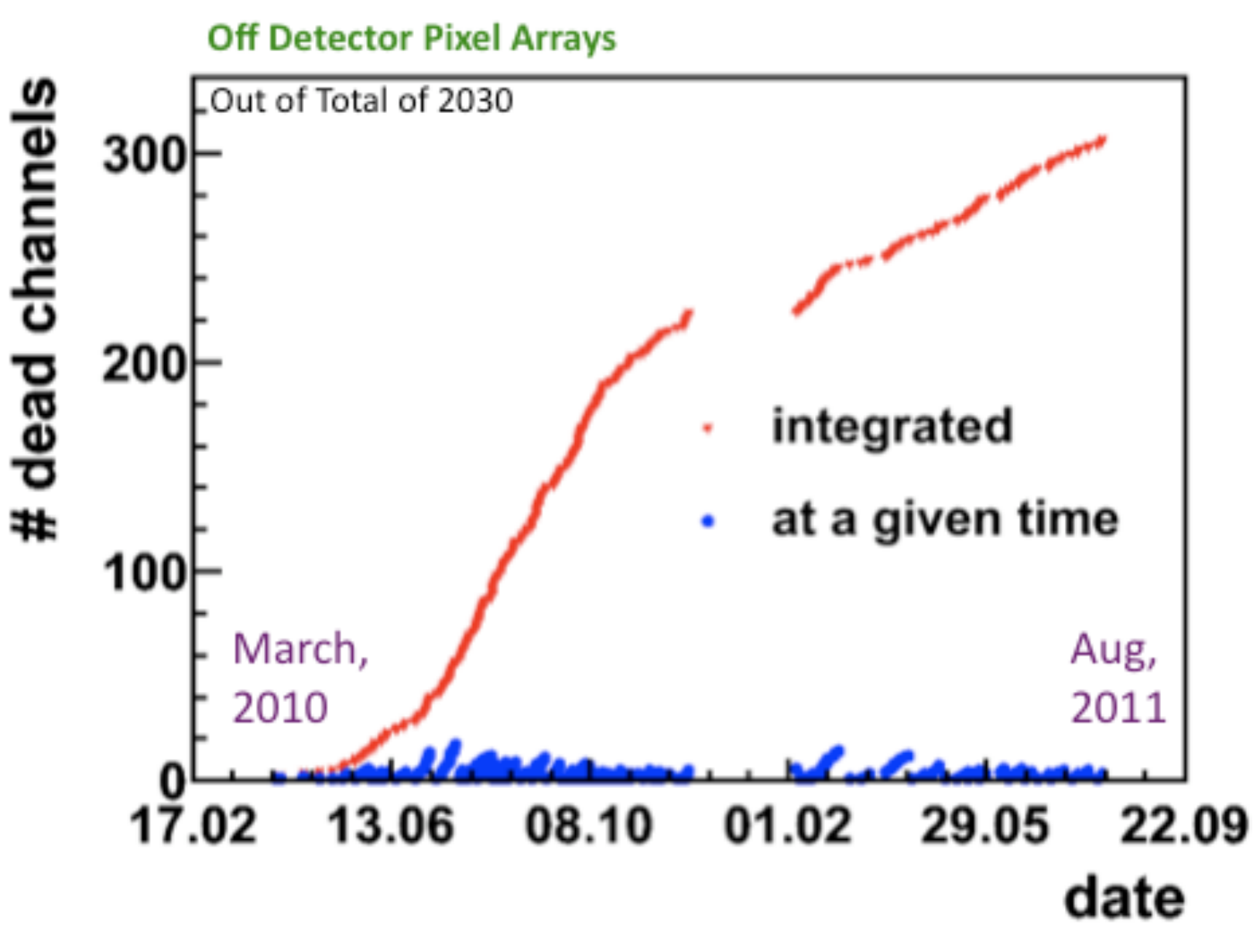}
%(a)
\end{minipage}
\hspace{0.5cm}
\begin{minipage}[b]{0.5\linewidth}
\centering
%(b)
\end{minipage}
\includegraphics[scale=0.46]{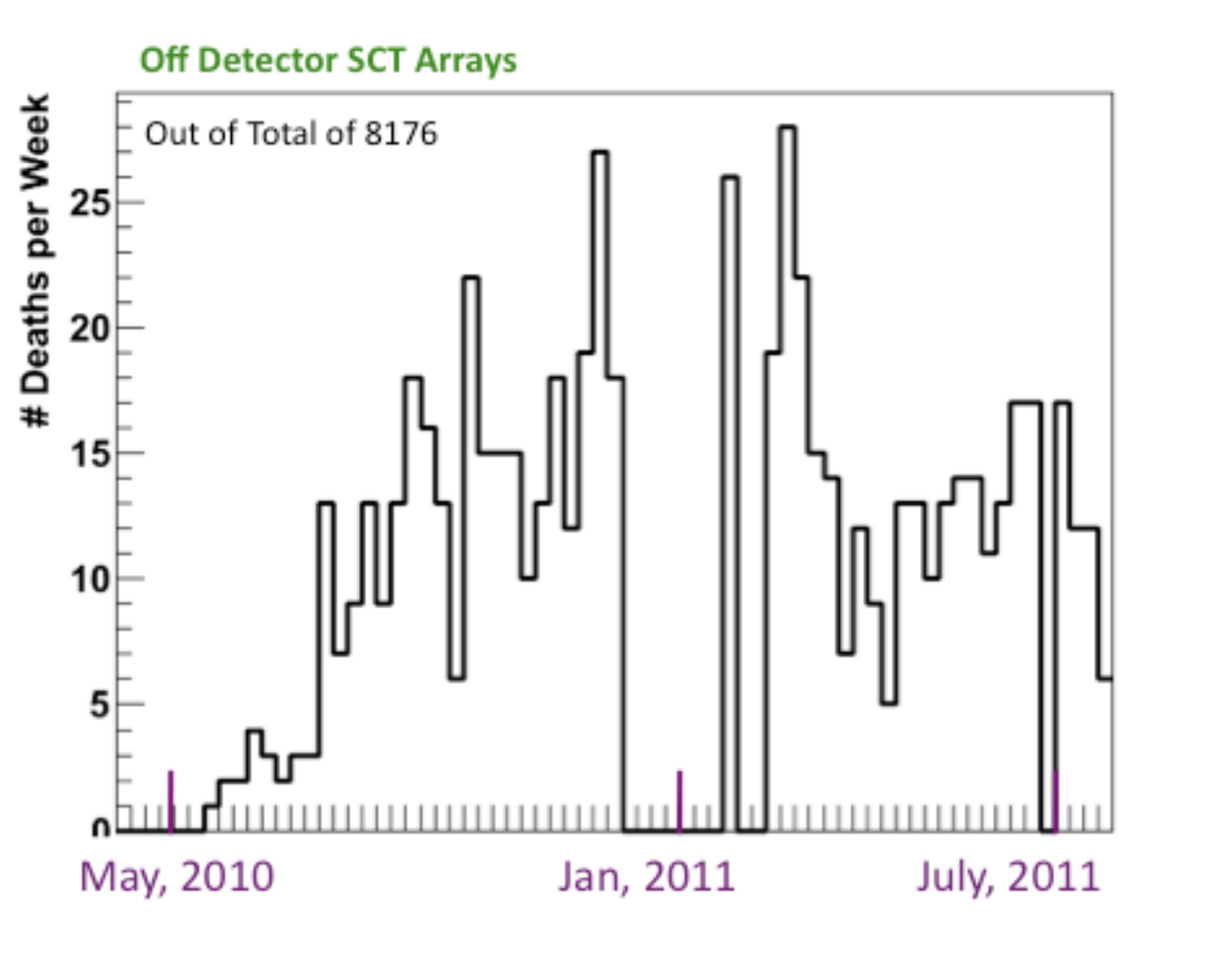}
\caption{ (Left)  The integrated number of failed pixel TX-plugin channels versus time. (Right) The number of SCT TX-plugin deaths per week versus time. }
\label{fig:failures}
\end{figure}

The pixel and SCT detectors have experienced VCSEL deaths in the off-detector optical readout electronics.  Early failures in the epoxy covered VCSEL arrays of the TX-plugins were observed in 2009, and a full replacement was performed with batches produced using a higher degree of quality assurance, particularly for ESD.  Nonetheless, failures continue to persist.  Figure~\ref{fig:failures}~(left) shows the integrated number of failed pixel TX-plugin channels versus time between March 2010 and August 2011.  Figure~\ref{fig:failures}~(right) shows the total number of SCT TX-plugin deaths per week versus time between May 2010 and July 2011.  In both cases, order 10 deaths occur per week.  As these failures occur in the off-detector electronics, they are continuously replaced during sufficiently long periods between LHC fills.

% --- Failures Figures ---
% -------------------------

%%%%%%%%%%%%%%%%%%%%%%%%%%%%%%%%%%
\section{Microscopic Analysis of a Failed Pixel Device}

\begin{figure}[tp]
\begin{minipage}[b]{0.5\linewidth}
\centering
\includegraphics[scale=0.40]{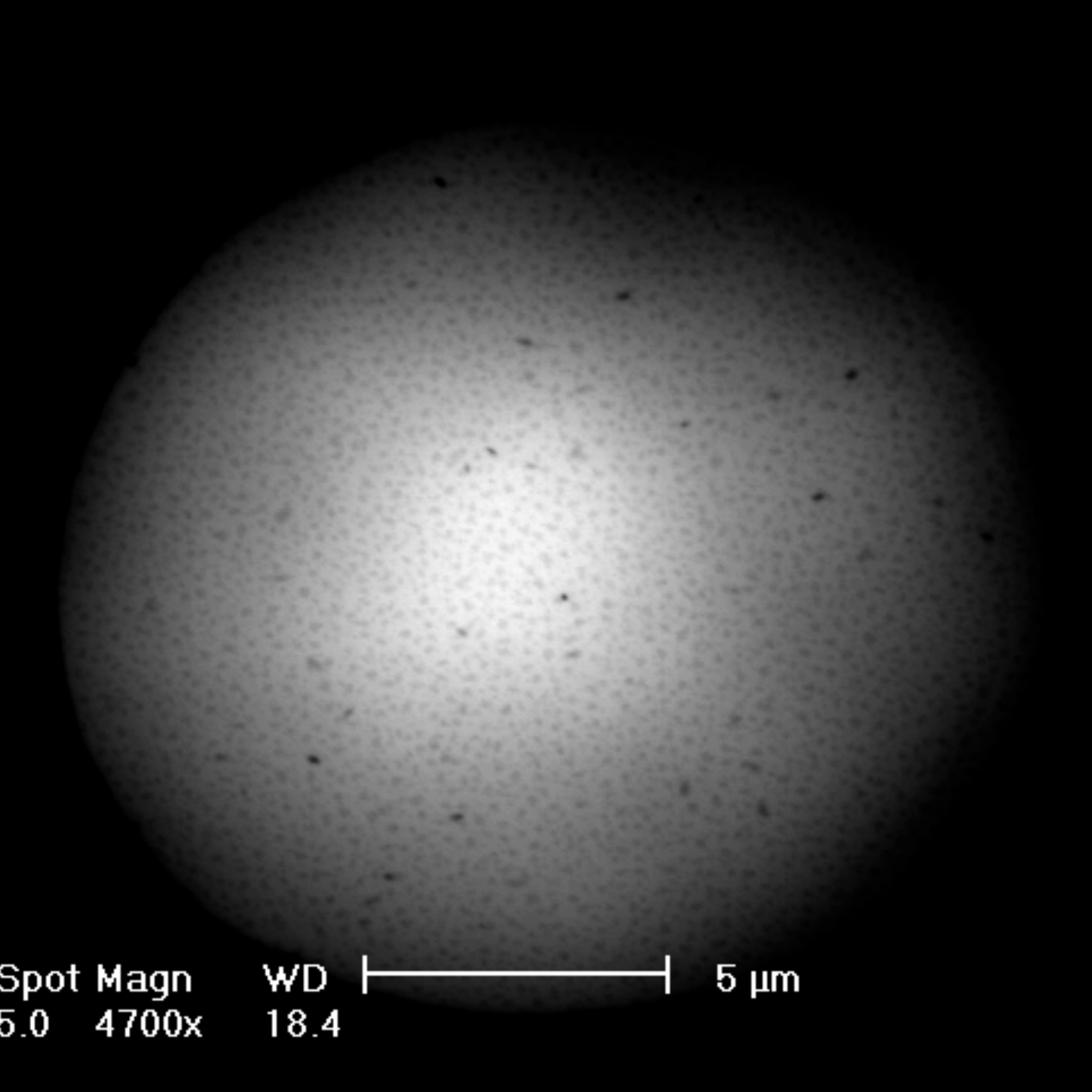}
%(a)
\end{minipage}
\hspace{0.5cm}
\begin{minipage}[b]{0.5\linewidth}
\centering
%(b)
\end{minipage}
\includegraphics[scale=0.40]{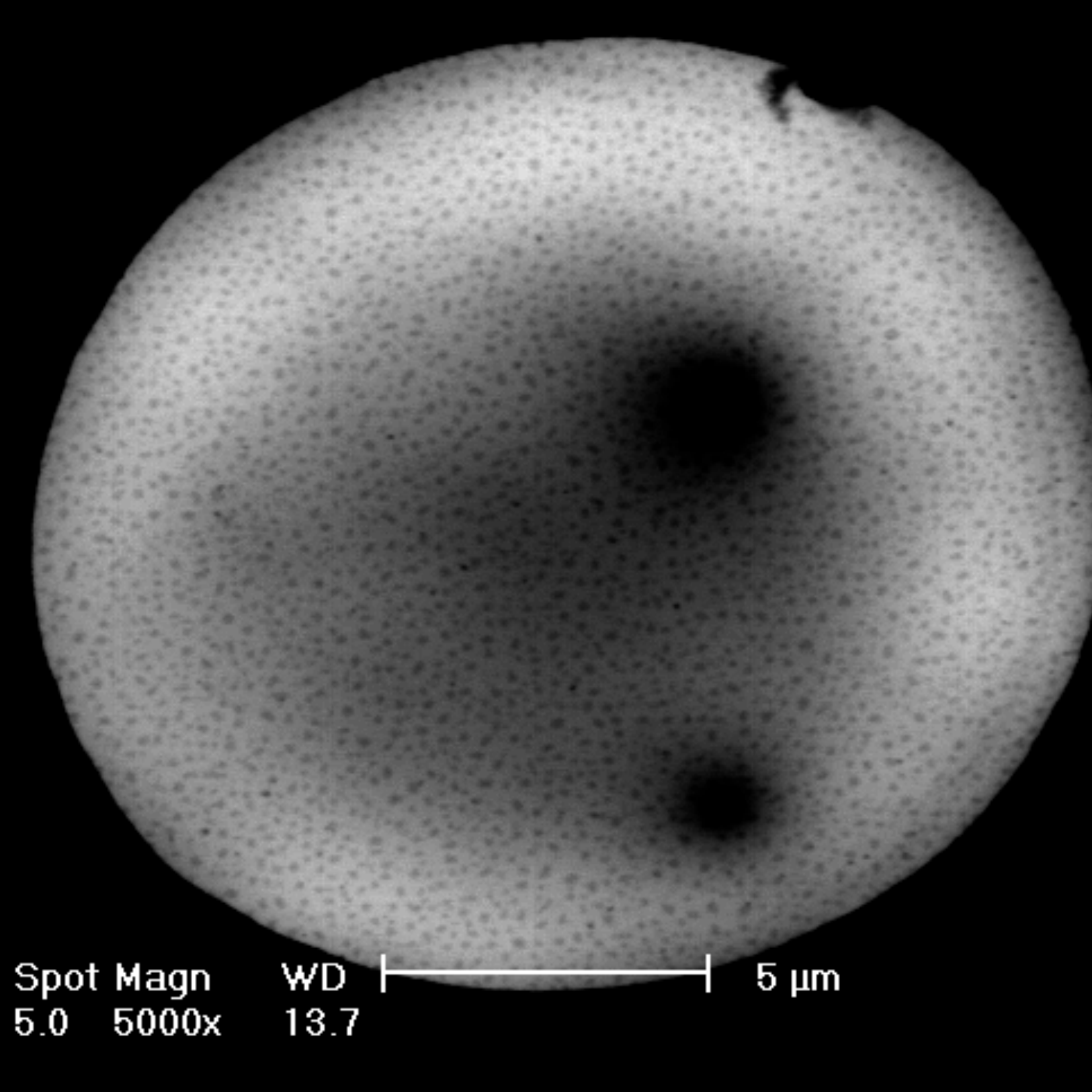}
\caption{ (Left) An EBIC image of a working VCSEL channel in a used pixel TX-plugin.  The small dark speckles reflect small various in surface topography. (Right) An EBIC image of a failed VCSEL channel in the same array as the channel shown on the left.  The image reveals two large defect regions.}
\label{fig:EBIC}
\end{figure}

Electron beam induced current (EBIC) analysis utilizes the scanning electron microscope beam to reveal defects in biased semiconductor devices, as the induced current will vary when the electron beam is scanned over such regions.  Figure~\ref{fig:EBIC}~(left) shows an EBIC image of a working VCSEL channel in a used pixel TX-plugin device.  The small dark speckles reflect small variations in surface topography.  Figure~\ref{fig:EBIC}~(right) shows an EBIC image of a failed VCSEL channel in the same array.  Two localized dark regions are evident.

% --- EBIC Figures ---
% -------------------------

\begin{figure}[tp]
\begin{minipage}[b]{0.5\linewidth}
\centering
\includegraphics[scale=0.45]{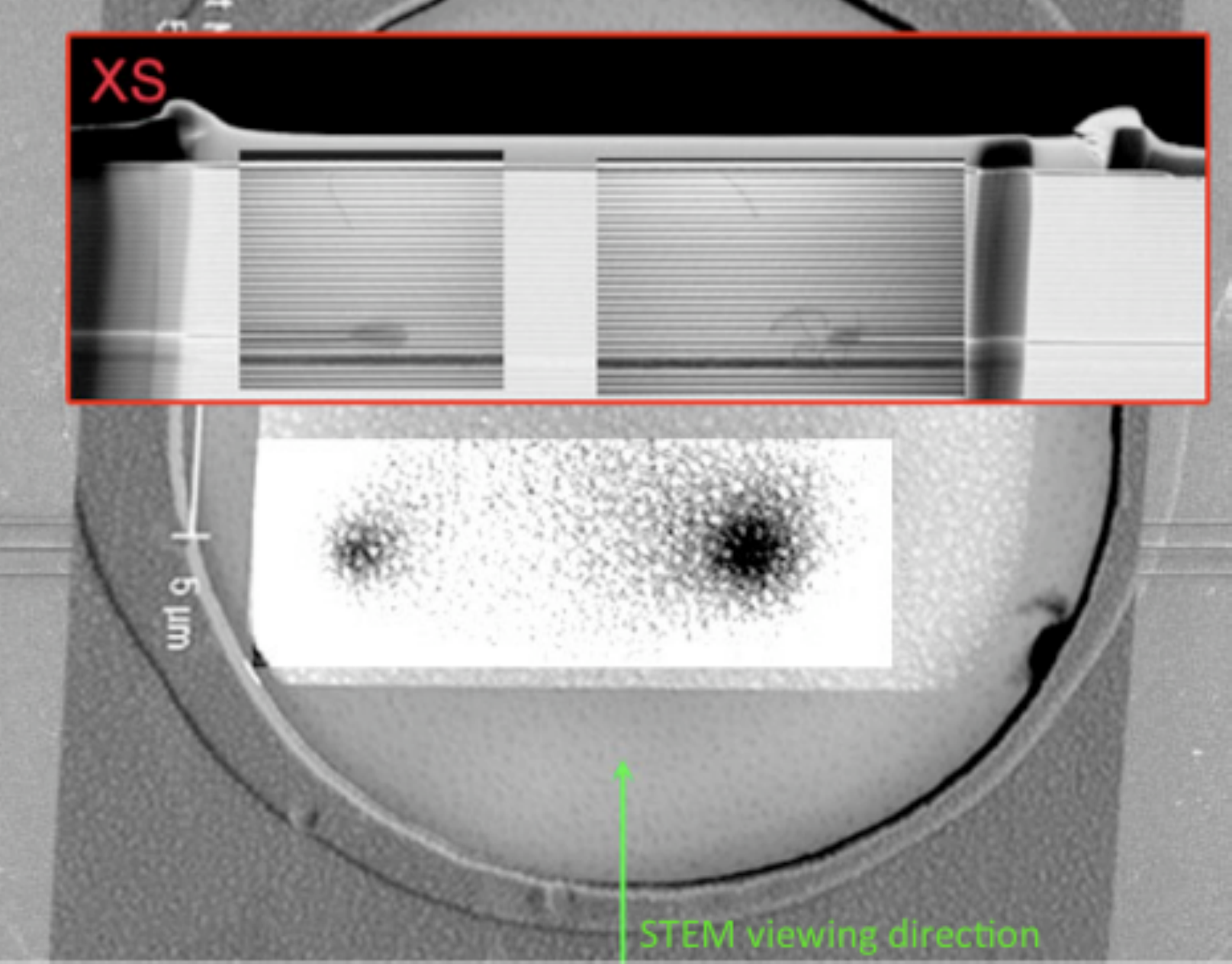}
%(a)
\end{minipage}
\hspace{0.5cm}
\begin{minipage}[b]{0.5\linewidth}
\centering
%(b)
\end{minipage}
\includegraphics[scale=0.45]{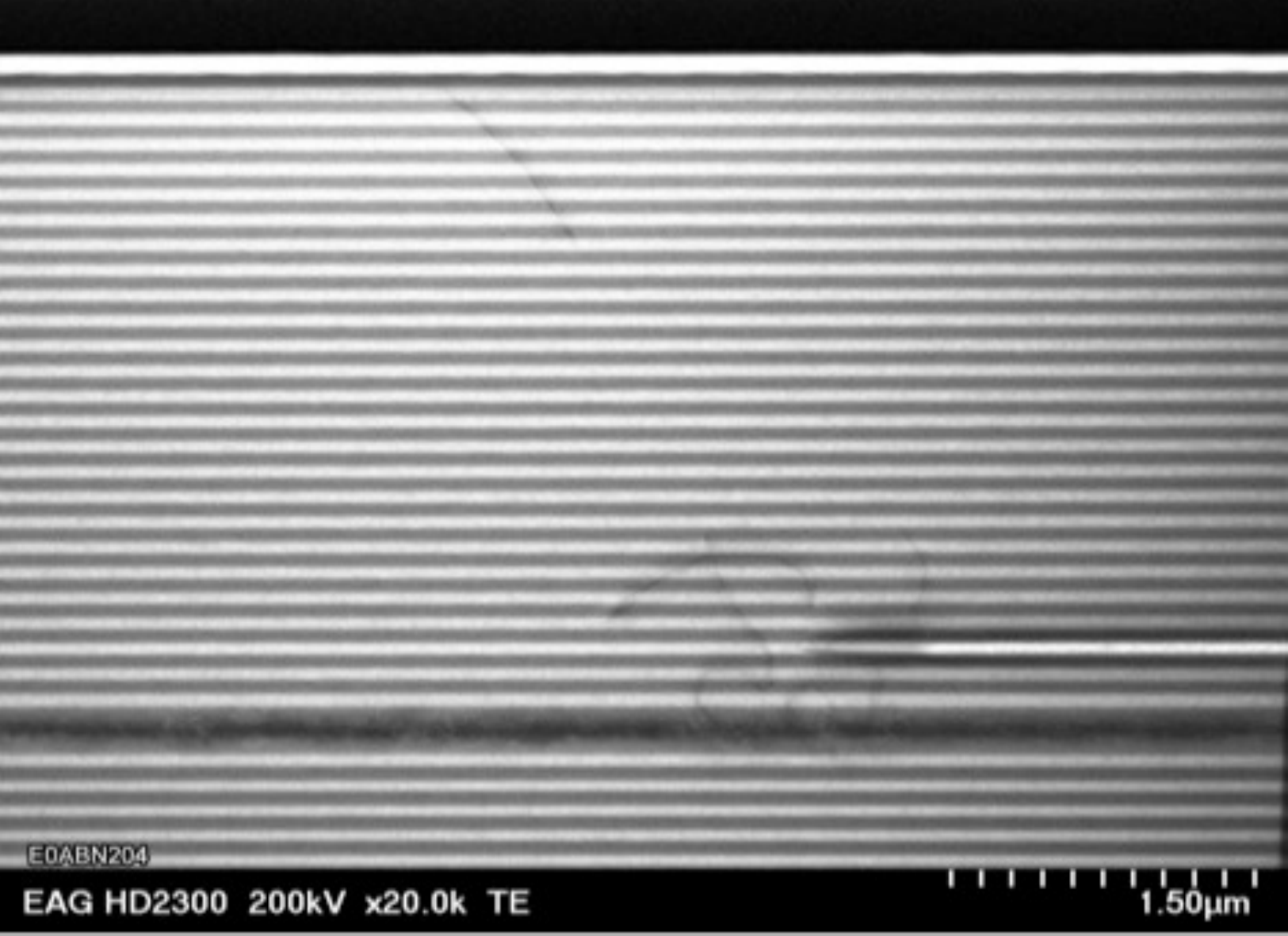}
\caption{ (Left)  A second EBIC plan view of the failed channel in Figure~\ref{fig:EBIC}~(right) indicating the region prepared for STEM analysis.  The boxed inlay labeled XS shows an STEM image of a cross-section slice through the defect regions.  (Right)  A zoomed-in STEM image of the right oxide layer aperture and neighboring quantum well region.}
\label{fig:STEM}
\end{figure}

The EBIC image of the failed VCSEL channel informed the preparation of a cross-sectional slice using a focused ion beam for analysis with a scanning transmission electron microscope (STEM).  Figure~\ref{fig:STEM}~(left) presents a second EBIC image of the failed VCSEL channel, and indicates the region prepared for STEM analysis.  The boxed inlay shows the STEM image of the cross section slice.  Dislocations are observed to emanate from the oxide layer, recognized as the white band protruding into the reflector region from the left and right of the inlay.  Figure~\ref{fig:STEM}~(right)  provides a zoomed view of the right oxide layer tip and the neighboring quantum well region.  In addition to the dislocation growth emanating from the oxide layer, defects are seen within the quantum well region, observed as the wide dark band layer.

\begin{figure}[tp]
\begin{minipage}[b]{0.5\linewidth}
\centering
\includegraphics[scale=0.45]{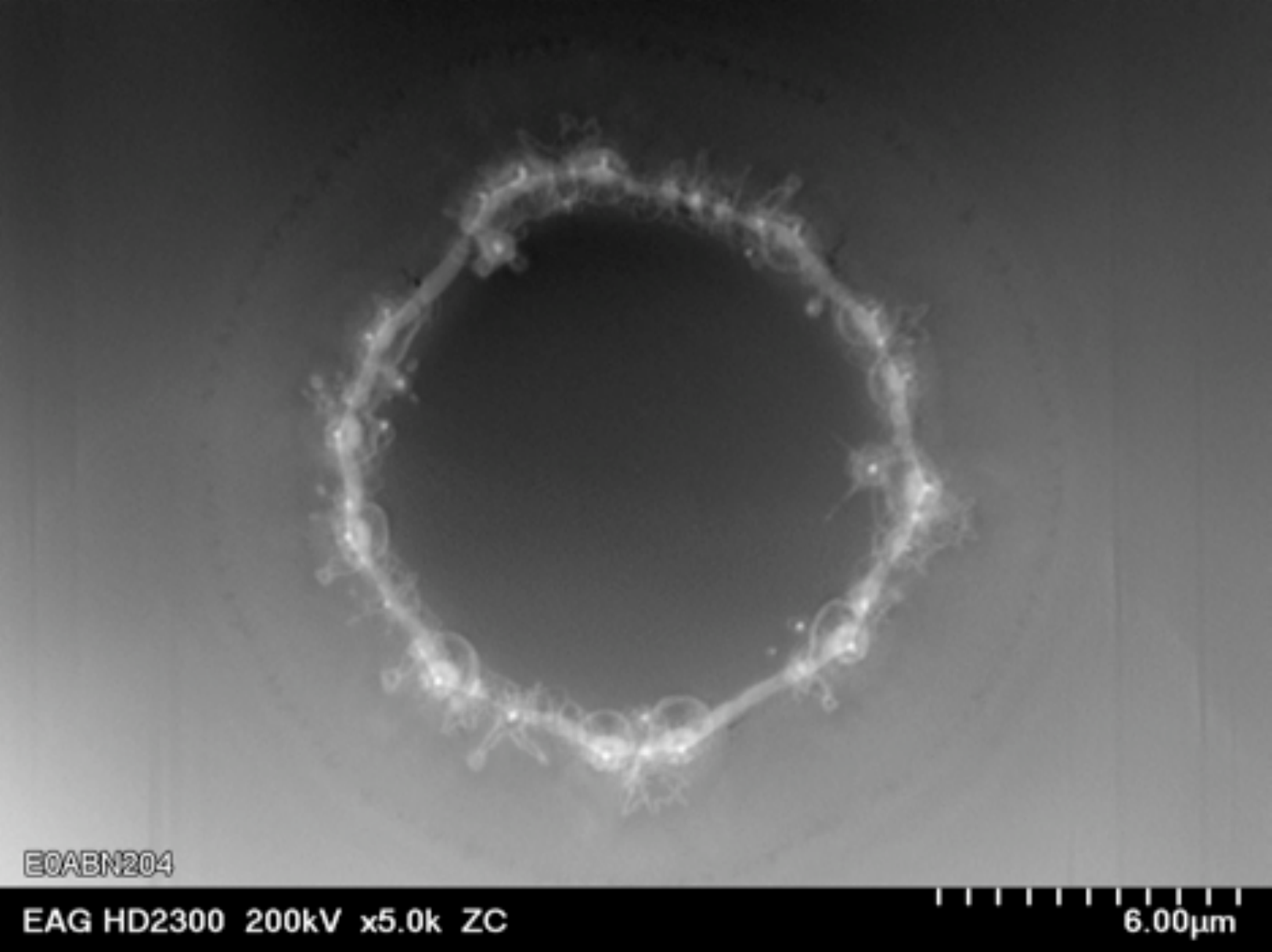}
%(a)
\end{minipage}
\hspace{0.5cm}
\begin{minipage}[b]{0.5\linewidth}
\centering
%(b)
\end{minipage}
\includegraphics[scale=0.45]{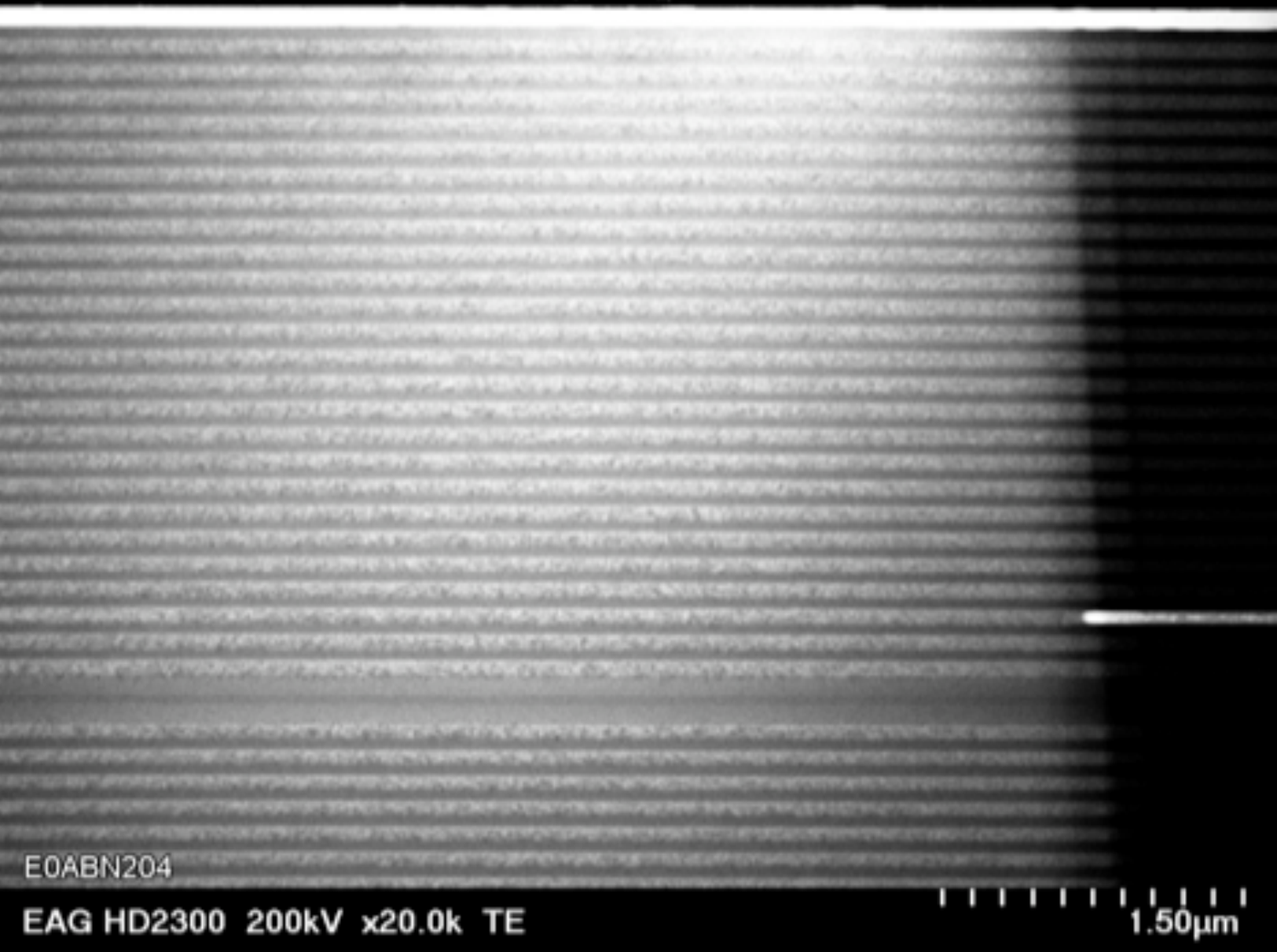}
\caption{ (Left) Plan view STEM image of a vertical slice of the working VCSEL channel containing the oxide and quantum well layers. (Right) Cross section STEM image of an unused VCSEL channel.}
\label{fig:STEM2}
\end{figure}

Figure~\ref{fig:STEM2}~(left) shows a plan view STEM image of a vertical slice containing the oxide and quantum well layers of the working channel whose EBIC image is shown in Figure~\ref{fig:EBIC}~(left).  A ring of dislocations about the oxide layer aperture is observed.  No such defects were seen on this channel in the EBIC analysis, as the defects were evidently too far below the surface.  The absence of dislocations within the ring suggests they had not yet grown into the active region and the channel therefore remained functional. 

The pixel array under investigation also contained an unbonded and hence unused VCSEL channel.  Figure~\ref{fig:STEM2}~(right) shows an STEM cross section image of this channel.  An absence of defects is clear when compared with the similar image of the failed device in Figure~\ref{fig:STEM}~(right).  

The microscopic analysis of a working, failed, and unused channel on a common array indicates that defects originating from the oxide layer during device operation propagate to the quantum well region, ultimately resulting in device failure.

% --- STEM (1) Figures ---
% -------------------------

% --- STEM (2) Figures ---
% -------------------------

%%%%%%%%%%%%%%%%%%%%%%%%%%%%%%%%%%
\section{Studies of the Stability of the Optical Spectrum Width}

The VCSELs in use by ATLAS are multi-mode devices.  Studies have shown that damage alters the properties of the lasing spectrum, typically resulting in a loss of optical modes and hence a narrowing of the spectral width~\cite{ESDref}.  The LAr collaboration measured the optical spectra of all on-detector OTx four times between 2009 and 2011~\cite{LARproc}.  The spectral widths were quite stable over the four measurements, and the majority of the devices displayed widths between 1 and 2.5~nm.  A sub-population of $\sim50$ VCSELs, however, had narrow widths between 0.5 and 1~nm.  All but one of the OTx failures which occurred after data on spectral widths were available were among the narrow width population.  During the replacement campaign in early 2011, all surviving narrow width OTx devices were preemptively replaced as well.  No new OTx deaths have occurred since this campaign.

As the LAr OTx are hermetically sealed, ESD damage at the manufacturer is a leading hypothesis for a cause of failure.  Another hypothesis is that the hermetic casing of a sub-population was compromised, allowing for electrolytic damage from humidity~\cite{HUMref}.  Recently, in a normal laboratory environment, the optical spectrum of a spare LAr OTx with a small hole drilled through the casing has been continuously measured over a period of $\sim100$ days.  The width of the spectrum has steadily narrowed during this test by $\sim1$~nm.

\begin{figure}[tp]
\begin{minipage}[b]{0.5\linewidth}
\centering
\includegraphics[scale=0.45]{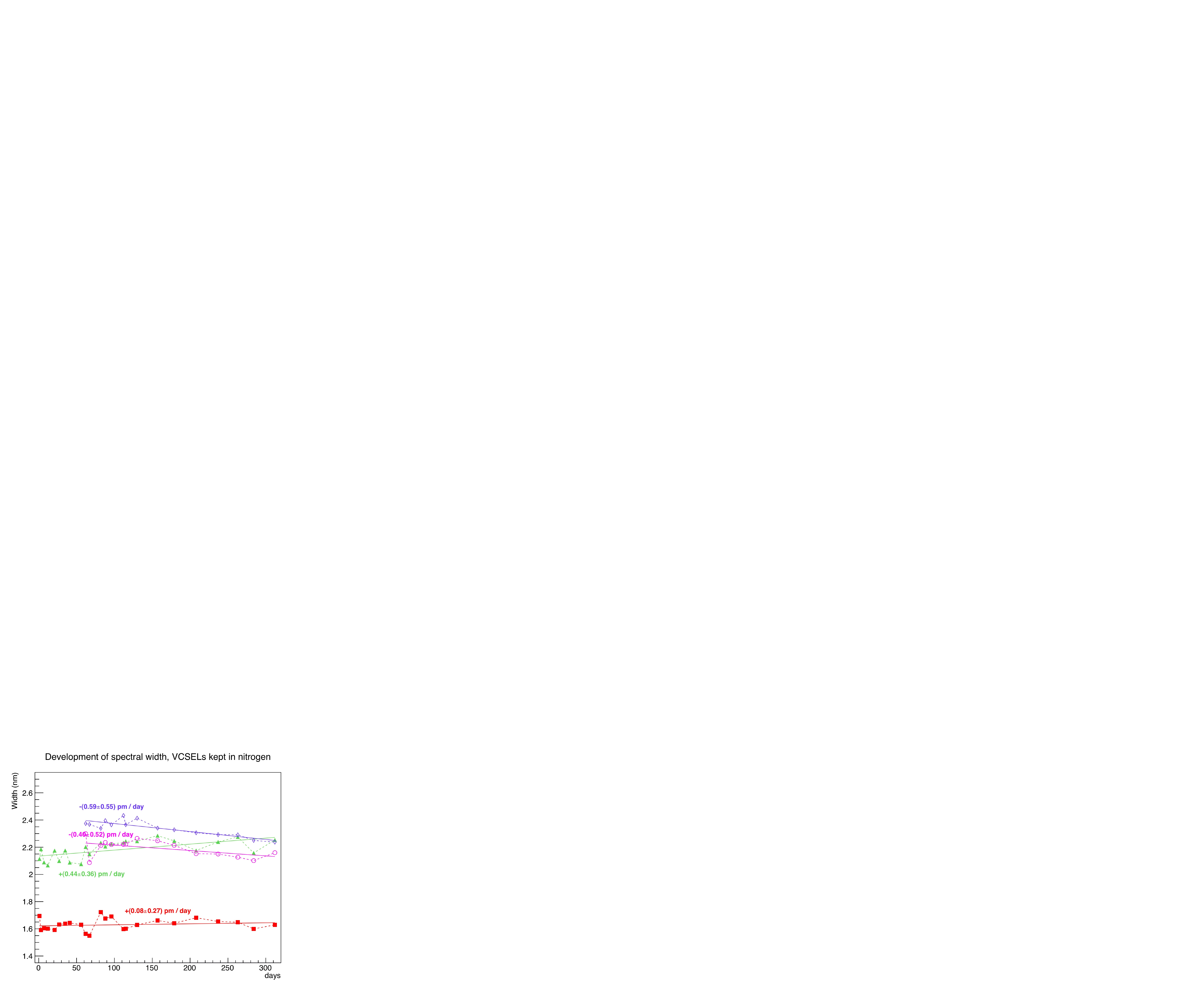}
%(a)
\end{minipage}
\hspace{0.5cm}
\begin{minipage}[b]{0.5\linewidth}
\centering
%(b)
\end{minipage}
\includegraphics[scale=0.45]{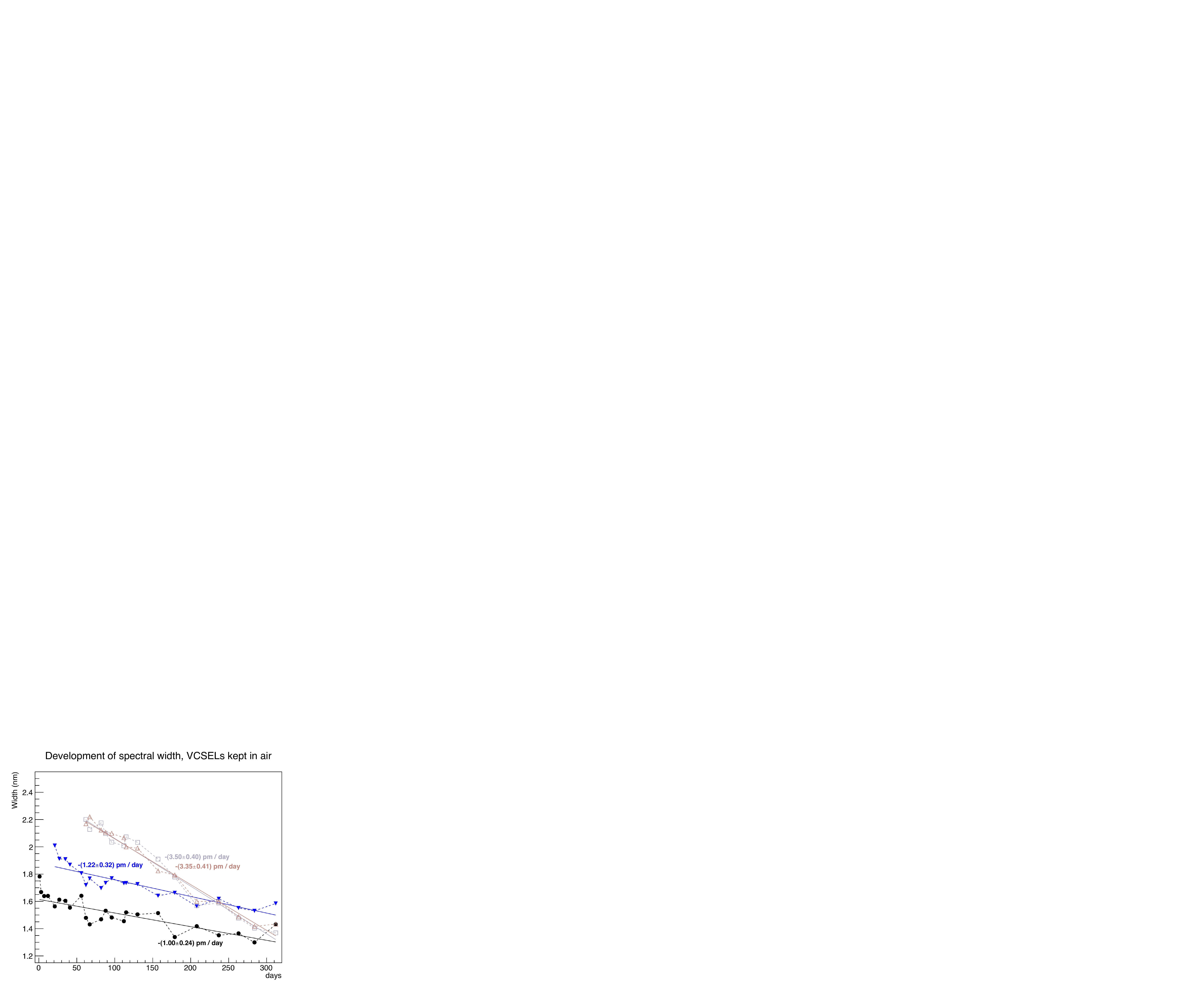}
\caption{(Left) Development of the spectral width over time of four TX-plugin arrays kept in nitrogen gas ($1\%$~relative humidity) at  $20^{\circ}\mathrm{C}$.  (Right)  Development of the spectral width over time of four TX-plugin arrays kept in air ($50\%$~relative humidity) at  $20^{\circ}\mathrm{C}$.}
\label{fig:SCTwidth}
\end{figure}

A similar test has also been performed by the SCT collaboration using spare TX-plugin arrays.  Figure~\ref{fig:SCTwidth}~(left) shows the average spectral width versus time of four spare TX-plugin arrays operated in a low humidity nitrogen gas environment at $20^{\circ}\mathrm{C}$.  The widths have remained stable for $\sim250$ days.  In contrast, Figure~\ref{fig:SCTwidth}~(right) shows the development of the width of four arrays operated in an air environment at $20^{\circ}\mathrm{C}$ and $50\%$ relative humidity.  Each array shows significant spectral narrowing in time when exposed to humid air.

%--- SCT Narrowing -------
%--------------------------------

%%%%%%%%%%%%%%%%%%%%%%%%%%%%%%%%%%
\section{Tests in Accelerated Lifetime Conditions}

In order to accelerate aging and study the time to device failure in a controlled environment, several tests have been carried out at higher temperatures than those discussed in the previous section, and in a greater variety of humidity conditions and device pre-history before operation.  Table~\ref{table:tests} summarizes the conditions and results of the tests, each of which was performed on approximately 30 TX-plugin VCSEL channels.

\begin{figure}[tp]
\begin{minipage}[t]{0.5\linewidth}
\centering
\includegraphics[scale=0.32]{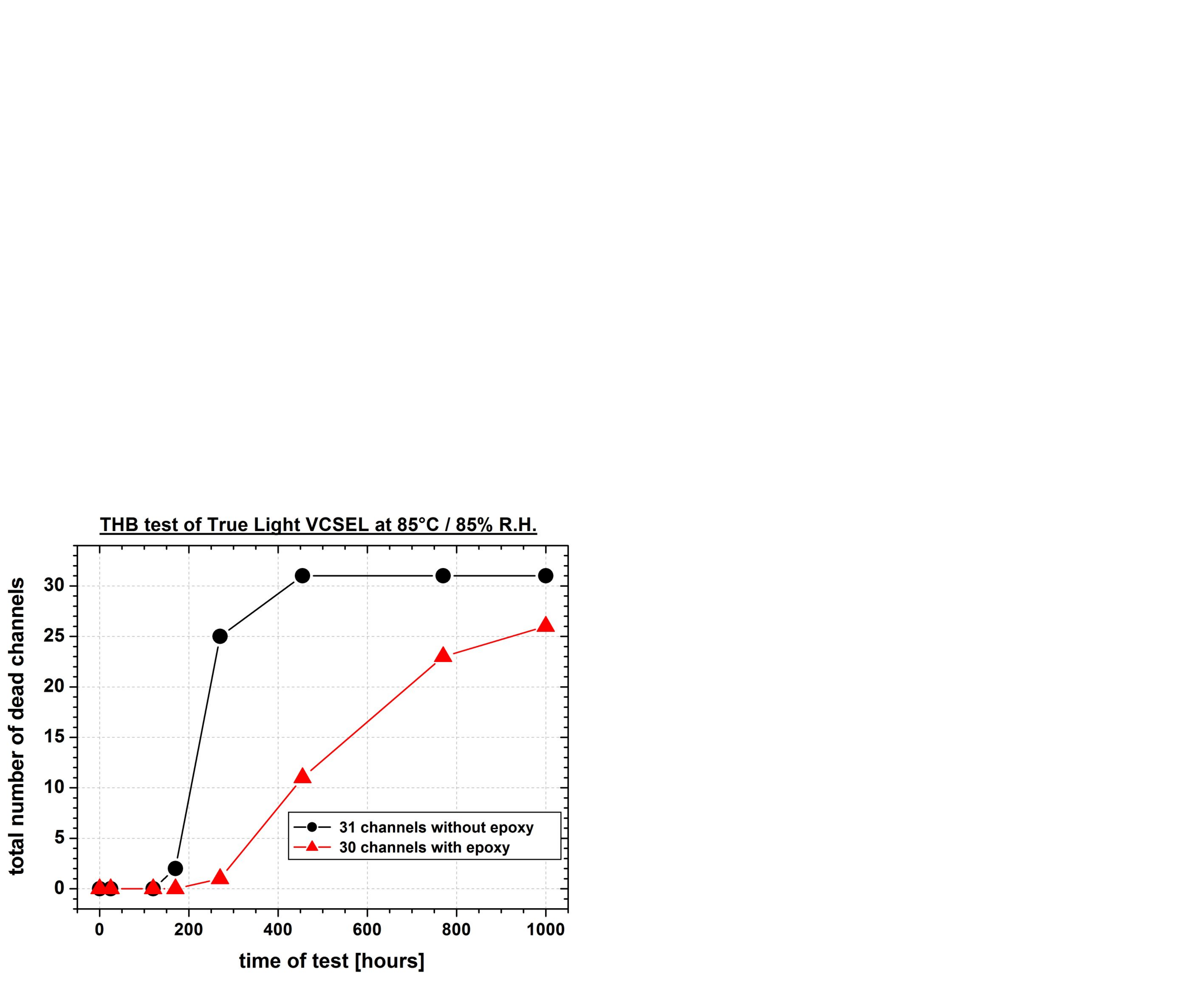}
%(a)
\end{minipage}
\hspace{0.5cm}
\begin{minipage}[t]{0.5\linewidth}
\centering
%(b)
\end{minipage}
\includegraphics[scale=0.32]{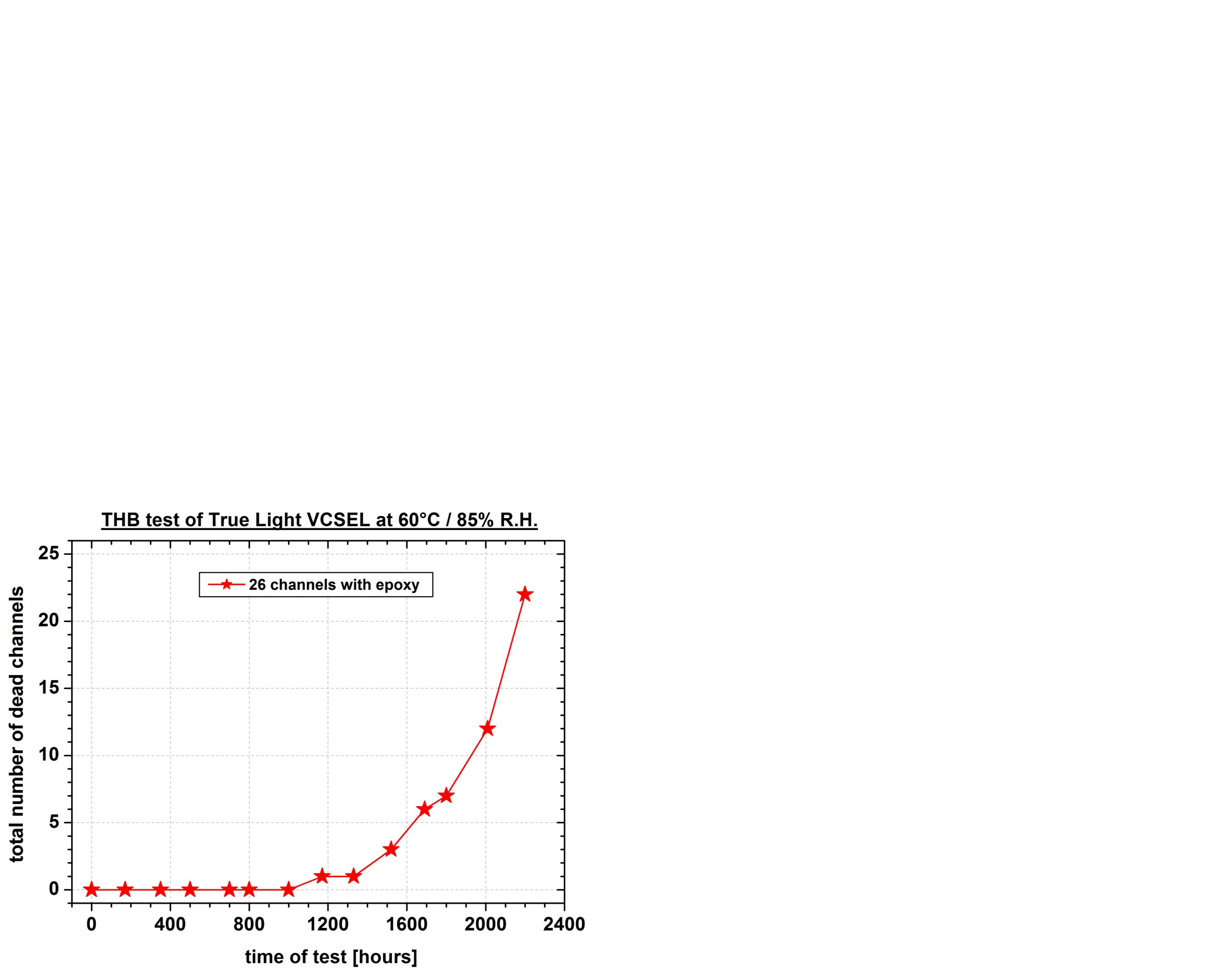}
\caption{(Left) The integrated number of failed VCSEL channels versus time in Test 1 at $85^{\circ}\mathrm{C}$ and 85\% relative humidity.  Arrays with and without epoxy coated were studied in this test.  (Right) The integrated number of failed VCSEL channels versus time in Test 2 at $60^{\circ}\mathrm{C}$ and 85\% relative humidity.  This test was performed on epoxy coated arrays.}
\label{fig:ULMresults}
\end{figure}

Test 1 was carried out at a temperature of $85^{\circ}\mathrm{C}$ and 85\% relative humidity, and examined arrays with and without epoxy coating.  The number of failed channels during the test is shown as a function of time in Figure~\ref{fig:ULMresults}~(left).  A device surviving operation in such conditions for 1000 hours is generally thought to be able to survive approximately 10 years in typical use conditions~\cite{HUMref}.  The channels on arrays without epoxy began to fail after only a couple hundred hours, and those on arrays with epoxy fared only marginally better.

Test 2 was performed with a lower temperature, $60^{\circ}\mathrm{C}$, and the same 85\% relative humidity as Test 1.  Only epoxy covered arrays were studied in this test.  Figure~\ref{fig:ULMresults}~(right) displays the results.  The first failures occurred shortly after 1000 hours, and nearly all had failed by 2200 hours.

% ----- Lifetime Test Table ----------
\begin{table*}[bp]
\begin{center}
\begin{tabular}{|c|c|c|c|c|c|c|}
\hline 
Test    & Temp. ($^{\circ}\mathrm{C}$) & Rel. Hum. (\%) & Abs. Hum. (g/m$^{3}$) & Notes           & Status & Result \\ \hline
1(a)     &       85        &       85                 &           363                        & w/ epoxy   & Complete  & 87\% dead after 1000 hrs \\
1(b)     &       85        &       85                 &           363                        & w/o epoxy & Complete  & 100\% dead after 1000 hrs \\ \hline
2          &       60        &       85                 &           125                        & w/ epoxy   & Complete  & 85\% dead after 2200 hrs \\ \hline
3(a)     &       85        &        0                  &              0                          & w/o epoxy, &Compete  & 0 deaths after 1650 hrs  \\  
            &                    &                            &                                           & pre-soaked 1k hrs 85\% R.H.  &   &  \\  
3(b)     &      85         &        0                  &              0                          &  w/ epoxy,                 &     In Progress  & 0 deaths after 1950 hrs \\ 
            &                    &                            &                                           & pre-soaked 1k hrs 85\% R.H., &                       &                                                 \\ 
            &                    &                            &                                           & then pre-biased 50 hrs, &                       &                                                 \\ 
            &                    &                            &                                           & then pre-baked 150 hrs  &   &    \\ \hline
4          &      85         &      29                  &           125                        &  w/ epoxy                  & In Progress &  0 deaths after 1090 hrs \\
\hline
\end{tabular}
\label{table:tests}
\caption{Summary of TX-Plugin VCSEL array accelerated lifetime tests under several extreme environmental conditions.}
\end{center}
\end{table*}
%--------------------------------------

The VCSEL failures observed in the LAr and inner detector readout systems have been oxide VCSELs operated under typical use conditions ($20-30^{\circ}\mathrm{C}$ and $30-50\%$ relative humidity).  Although to date there is not a similar failure issue with the on-detector oxide VCSEL arrays in the pixel opto-boards, the possibility remains a concern, as they are of the same make as the original off-detector batch and would not be accessible for replacement until a prolonged LHC shutdown period.  There are, however, several noteworthy differences between the on and off-detector pixel VCSELs.  First, the opto-board VCSELs are only significantly biased when sending data for triggered events, while the off-detector VCSELs transmit the 40 MHz clock to the modules.  Further, the opto-board VCSELs are operated at $18^{\circ}\mathrm{C}$ in a dry nitrogen gas environment within the inner detector envelope.  

The third set of tests study the lifetime at $85^{\circ}\mathrm{C}$, as in Test 1, but in a very low relative humidity environment (essentially 0\%).  Test 3(b), in particular, also attempts to closely emulate the pre-history of the pixel opto-boards, which includes possible exposure to humidity and bias during system integration and testing before the detector was installed and operated in the dry inner detector environment.  No failures have occurred in the first 2000 hours of this test.

With the data from the accelerated lifetime tests, and the failure data of the devices in the detector readout systems, analyses are currently in progress which attempt to extrapolate between the measurements using various acceleration models.  For example, the well known Arrhenius model predicts an acceleration factor which is dependent on the temperature.  Humidity is evidently also a relevant variable in determining the time to failure.  An open question is whether the relevant formulation of the humidity dependence is in terms of the relative or absolute humidity.  Test 4, performed at  $85^{\circ}\mathrm{C}$ and $29\%$ relative humidity, is designed to address this question.  The absolute humidity humidity in Tests 2 and 4 have the same value, 125 g/m$^{3}$.

%%%%%%%%%%%%%%%%%%%%%%%%%%%%%%%%%%
\section{Contingency Plans}

Each detector system is developing contingency plans.  The LAr collaboration is developing new OTx packages~\cite{LARproc}, which would replace the existing ones and be installed during the 2013 LHC shutdown.  One of the OTx prototypes utilizes redundant optical links.   Likewise, the pixel and SCT collaborations are studying more robust VCSEL arrays for TX-plugins.  VCSELs from two manufactures, AOC and ULM, have been developed to be humidity resistant.  The AOC arrays have been subjected to accelerated lifetime tests with excellent results.  Similar tests are planned for the ULM VCSELs.  Lastly, a pixel collaboration project is currently underway to fabricate new service panels which would allow the on-detector VCSELs to be installed in a more accessible region of the detector.  If needed, this new optical readout system would be installed during the 2013 LHC shutdown.

%%%%%%%%%%%%%%%%%%%%%%%%%%%%%%%%%%
\section{Conclusions}

The ATLAS LAr calorimeter and silicon tracking detectors employ oxide VCSEL based optical readout systems, and these devices have failed well before their expected lifetime.  An analysis of a failed device shows evidence of oxide layer damage, with dislocations growing into the active layer.  ESD and humidity induced corrosion are known to cause this type of damage.  Narrowing of the width of the optical spectrum is being studied as an indicator of device damage, and tests in high temperature and variable humidity conditions are being performed to validate models which predict the time to failure.  While these studies are being performed, contingency plans are being executed in parallel.  In particular, significant modifications to the LAr and pixel readout systems may be introduced during the 2013 LHC shutdown.

%%%%%%%%%%%%%%%%%%%%%%%%%%%%%%%%%%
\begin{acknowledgments}
This work has been carried out by several collaborating institutions including Academia Sinica, Bergische Universit\"{a}t Wuppertal, Columbia University Nevis Laboratories, Laboratoire de l'Acc\'{e}l\'{e}rateur Lin\'{e}aire d'Orsay (IN2P3-LAL), Lawrence Berkeley National Laboratory, The Ohio State University, Organization Eurp\'{e}ene pour la Recherche Nucl\'{e}aire (CERN), Oxford University, STFC Rutherford Appleton Laboratory, Southern Methodist University, University of California, Santa Cruz, Universit\'{e} de Gen\`{e}ve, and Universit\"{a}t Siegen.  We also acknowledge the services of Evans Analytical Group and ULM Photonics.
\end{acknowledgments}

\bigskip % extra skip inserted
%%%%%%%%%%%%%%%%%%%%%%%%%%%%%%%%%%

\end{document}